\documentclass[twocolumn,showpacs,showkeys,superscriptaddress]{revtex4}
\usepackage{graphicx,subfigure,epsfig}
\usepackage{amsfonts}
\usepackage{graphicx}
\usepackage{amsmath}

\newcommand{\bastar}{\begin{eqnarray*}}
\newcommand{\eastar}{\end{eqnarray*}}
\newskip\humongous \humongous=0pt plus 1000pt minus 1000pt

\newif\ifdtup

\relax
\newcommand{\bea}{\begin{eqnarray}}
\newcommand{\eea}{\end{eqnarray}}

\begin{document}
\title {Solution-generating methods of Einstein's equations 
by Hamiltonian reduction}

\author{Seung Hun Oh}
\email{shoh.physics@gmail.com}
\affiliation{Department of Physics, Konkuk University,  Seoul 05029, Korea}

\author{Kyoungtae Kimm}
\affiliation{Faculty of Liberal Education,
Seoul National University, Seoul 08826, Korea}

\author{Yongmin Cho}
\affiliation{Administration Building 310-4 
Konkuk University, Seoul 05029, Korea}
\affiliation{School of Physics and Astronomy 
Seoul National University,
Seoul 08826, Korea}

\author{Jong Hyuk Yoon}
\email{yoonjh@konkuk.ac.kr}
\affiliation{Department of Physics, Konkuk University,  Seoul 05029, Korea}

\begin{abstract}
The purpose of this paper is to demonstrate a new method of generating exact solutions to the Einstein's equations obtained by the Hamiltonian reduction.
The key element to the successful Hamiltonian reduction is finding
the privileged spacetime coordinates in which physical degrees of freedom manifestly reside in the conformal two-metric, and all the other metric components are determined by the conformal two-metric. In the privileged coordinates the Einstein's constraint equations become 
{\it trivial}; the Hamiltonian and momentum constraints are simply the defining equations of a non-vanishing gravitational Hamiltonian and momentum densities in terms of conformal two-metric and its conjugate momentum, respectively. Thus, given any conformal two-metric, 
which is a constraint-free data, one can construct the whole 4-dimensional spacetime by integrating the first-order {\it superpotential} equations.
As the first examples of using Hamiltonian reduction in solving the Einstein's equations, 
we found two exact solutions to the Einstein's equations in the privileged coordinates. Suitable coordinate transformations from the privileged to the standard coordinates show that
they are just the Einstein-Rosen wave and the Schwarzschild solution. The local gravitational Hamiltonian and momentum densities of these spacetimes are also presented in the privileged coordinates.
\end{abstract}
\maketitle

\section{Introduction}
As is well-known, general relativity (GR) is a parametrized theory in the sense  that the Hamiltonian and momentum densities of gravitational fields are pure  constraints \cite{Dirac, ADM}.  This is due to the presence of arbitrariness in slicing a given spacetime into a family of 3-dimensional spacelike hypersurfaces and the freedom to introduce any coordinates on each of them as one wishes. Indeed, these constraints form the algebra of arbitrary deformations of a given spacetime \cite{KT}. If one could choose certain functions of gravitational phase space as privileged spacetime coordinates such that the Hamiltonian and momentum constraints could be solved {\it trivially}, then one could well expect to find a non-zero local Hamiltonian and momentum densities of GR in these privileged coordinates.
This procedure is known as Hamiltonian reduction first introduced by Arnowitt, Deser, and Misner (ADM) using the (3+1) decomposition of a 4-dimensional spacetime \cite{ADM},
but with only a partial success.
Later, Prof. K. Kucha{\u r} studied the Hamiltonian reduction of GR, assuming two commuting Killing vector fields, and showed that Einstein's theory is identical to the field theory of a free massless scalar field propagating in the Minkowski spacetime \cite{kuchar71}.

The standard ADM Hamiltonian reduction based on the (3+1) decomposition \cite{ADM,kuchar71,fischer97,fischer00}, however, is not the only way of realizing Hamiltonian reduction of GR, because a 4-dimensional spacetime could be also viewed as a local product of a 2-dimensional base manifold with the Lorentzian signature and a 2-dimensional spacelike manifold \cite{cho92}.
Based on the (2+2) decomposition \cite{dinverno78,Yoon0}, one of us recently proposed a new scheme of carrying out the Hamiltonian reduction without assuming isometries \cite{Yoon1, Yoon2}. 
Specifically, the area of the 2-dimensional cross-section of the out-going null hypersurface is chosen as the physical time coordinate $\tau$, and the radial coordinate $R$ is defined by ``equi-potential'' function in the gravitational phase space, and the remaining two spatial coordinates $Y^{a} (a= 1,2)$ are introduced on each  ``equi-potential'' surface such that the {\it two-dimensional shift} is zero.

The (2+2) Hamiltonian reduction in the privileged coordinates $(\tau, R, Y^{a})$ has the following remarkable features. The Einstein constraint equations turn out to be merely the equations 
that define a non-vanishing local Hamiltonian and momentum densities of gravitational fields. Namely, 
the Hamiltonian constraint is {\it trivially} solved to define a non-vanishing gravitational Hamiltonian density as a function of the conformal two-metric of the null hypersurface and its conjugate momentum, which dictates the evolutions in the physical time of the canonical pair of the conformal two-metric and its conjugate. The momentum constraints are also solved {\it trivially} to define non-vanishing gravitational momentum densities carried by the conformal two geometry, which assume the canonical form of local momentum densities in standard field theories.
True physical degrees of freedom of gravitational fields reside in the conformal two-metric, with all other metric components determined by the conformal two-metric.
Therefore, solving the Hamilton's equations of motion of the conformal two-metric and its conjugate momentum is a crucial step in the process of solving the Einstein's equations in the (2+2) Hamiltonian reduction. In addition, one of the Einstein's equations shows up as a topological constraint equation that restricts the spatial topology of the cross-section of 
the null hypersurface either as a two-sphere or a torus.

The main purpose of this paper is to illustrate a new method of generating exact solutions to
the Einstein's equations using the (2+2) Hamiltonian reduction. This method consists of the following procedure. Firstly, one has to determine the $\tau$ and $R$ dependence of the conformal two-metric and its conjugate momentum by solving their evolution equations.
Secondly, one must solve the topological constraint equation in order to determine 
the $Y^{a}$ dependence of the conformal two-metric and its conjugate momentum. 
It is unnecessary to solve the Einstein's constraint equations, as they are {\it trivial}. They are simply defining equations of the gravitational Hamiltonian and momentum densities in terms of the conformal two-metric and its conjugate momentum. Thirdly, one has to find the 
{\it superpotential} by integrating the first-order differential equations, which are again determined by the conformal two-metric and its conjugate momentum via the Hamiltonian and momentum densities. The remaining Einstein's equations are either trivial identities or redundant equations. This completely determines the spacetime metric in the privileged coordinates $(\tau, R, Y^{a})$.

Following the above procedure, we generate non-trivial solutions to the Einstein's equations in the privileged 
coordinates $(\tau, R, Y^{a})$, and identify them as the Einstein-Rosen waves and the Schwarzschild solution after making suitable coordinate transformations.
This paper is organized as follows. In Section 2, we will review the main results of the (2+2) Hamiltonian reduction and present the Einstein's equations in the privileged coordinates.
In Sections 3, we will determine the conformal two-metric and its conjugate momentum in the privileged coordinates using an ansatz. Then we integrate the superpotential equations to determine the whole spacetime in a way consistent with the topological constraint equation. 
We show that this solution is identical to the Einstein-Rosen spacetime by making suitable coordinate transformations to the standard coordinates.
In Section 4, we solve the the conformal two-metric and its conjugate momentum after making
a {\it complex} coordinate transformation of the privileged coordinates, and determine the
superpotential and the whole spacetime metric. It turns out that this solution is 
the Schwarzschild metric in the canonical Weyl coordinates. For an illustration purpose, we also present the Hamiltonian and momentum densities of the Schwarzschild metric in the privileged coordinates. 
In Section 5, we summarize and discuss possible generalizations of the (2+2) Hamiltonian reduction to include a non-vanishing two-dimensional shift vector.

\section{Review of the (2+2) Hamiltonian Reduction}
Geometrically, any 4-dimensional spacetime can be regarded as a fibre bundle $E_4$ that consists of a 2-dimensional Lorentzian base manifold and a 2-dimensional fibre $N_2$ at each point of the base manifold. In this paper,
we restrict our considerations to the case where the topology of spacelike 
hypersurface of $E_4$ is $\mathbb{R} \times N_2$. Then,
we may introduce the privileged coordinates $(\tau , R , Y^1 , Y^2)$ on $E_4$, with the line element given by \cite{Yoon1,Yoon2}
\bea
ds^2 = -2h (2 dR d\tau + dR^2 ) + \tau \rho_{ab}dY^a dY^b ,  \label{ohyoonmetric}
\eea
where $a,b = 1,2$. Here, $\rho_{ab}$ is the conformal two-metric on $N_2$ 
with a unit determinant and $\tau$ is the area element of $N_2$. 
We choose the sign $h<0$ so that  $\tau=$ constant hypersurface is spacelike. Thus, the area element 
$\tau$ plays the role of time, hence the name the \textit{area time}. When no isometries are
present, the functions $h$ and $\rho_{ab}$ that appear in the metric
(\ref{ohyoonmetric}) depend on all the coordinates $(\tau , R , Y^1 , Y^2)$.

In the (2+2) Hamiltonian reduction of GR, it was shown that the Einstein's equations can be
written as the following five sets of equations (i), (ii), $\cdots$, (vi) in the privileged coordinates \cite{Yoon1,Yoon2}:

\noindent
(i) Four constraint equations are solved to define the local Hamiltonian density $-\pi_{\tau}$, local momentum densities $\pi_{R}$, and $\tau^{-1}\pi_{a}$ as follows,
\bea
& &  -\pi_{\tau} = {\mathcal H} - 2 \partial_{R}\ln (-h),                       
 \label{pitau}\\
& & \pi_{R}=-\pi^{ab}\partial_{R}\rho_{ab},
\label{radi} \\
& & \tau^{-1}{\pi}_{a}
= - \pi^{bc}{\partial \over \partial Y^{a}} \rho_{bc}  
+2{\partial \over \partial Y^{b}}(\pi^{bc}\rho_{ac})                \nonumber\\ 
& & \hspace{0.55in}
- {\partial \over \partial Y^{a}}\{ \tau ( {\mathcal H} + \pi_{R})\},  \label{angko}
\eea
where $\mathcal{H}$ is given by
\bea
& & 
{\mathcal H}= \tau^{-1} \rho_{a b}\rho_{c d}\pi^{a c}\pi^{b d}
+{1\over 4}\tau \rho^{a b} \rho^{c d}
(\partial_{R}\rho_{a c}) (\partial_{R}\rho_{b d})  \nonumber \\
& & \hspace{0.3in}
+\pi^{a c}\partial_{R}\rho_{a c} +  {1\over 2\tau},
\label{nail}     
\eea
and $\pi^{a b}$ is the traceless momentum conjugate to the conformal two-metric 
$\rho_{a b}$  ($\rho_{ab} \pi^{ab} = 0$).

\noindent
(ii) There are four equations that relate the gradient of the {\it superpotential}
 $\ln (-h)$ to ${\mathcal H}$, $\pi_{R}$, and $\tau^{-1}\pi_{a}$ as follows,
\bea
& & \partial_{\tau}  {\ln}(-h) = {\mathcal H} - \tau^{-1}, \label{sp1} \\
& & \partial_{R}  {\ln}(-h) = -\pi_{R}, \label{sp2} \\
& & \partial_a {\ln}(-h) = -\tau^{-1}\pi_{a}. \label{sp3}
\eea
\noindent
(iii) The integrability conditions follow trivially from the above equations (\ref{sp1}), (\ref{sp2}), and (\ref{sp3}),
\bea
& & {\partial \pi_{R} \over \partial \tau} 
= - {\partial {\mathcal H} \over \partial R}, \label{patikim}\\
& & {\partial \over \partial \tau} (\tau^{-1}\pi_{a})
=-{\partial \over \partial Y^{a}} {\mathcal H},        \label{sycamore}\\
& & {\partial  \over \partial R}(\tau^{-1}\pi_{a})
={\partial \over \partial Y^{a}}\pi_{R}.    \label{limb}
\eea

\noindent
(iv) 
There is the \textit{topological constraint equation} \cite{lewandowski04}
\begin{equation}
 R^{(2)}_{ab} -{1\over 2}\tau^{-2}\pi_{a}\pi_{b} 
 +\nabla^{(2)}_{a}  (\tau^{-1}\pi_{b}) =0,    \label{topological}
\end{equation}
where $R^{(2)}_{ab}$ is the Ricci tensor of $N_{2}$, and $\nabla^{(2)}_{a}$ is the covariant derivative on $N_{2}$. The equation (\ref{topological}) determines the $Y^{a}$ dependence of the conformal two-metric. The trace
of the above equation is 
\begin{equation}
\tau R^{(2)} -{1\over 2}\tau^{-2}\rho^{ab}\pi_{a}
\pi_{b} + { \partial \over \partial Y^{a}} (\tau^{-1}\rho^{ab}\pi_{b}) =0.
\label{CS}
\end{equation}
For a \textit{compact} $N_2$, the integral of the equation (\ref{CS}) over $N_2$ becomes
\bea
\int_{N_{2}}\!\!\!\! d^{2}Y \tau^{-2}\rho^{ab}\pi_{a}\pi_{b}
= 16\pi (1-g) \geq 0,    \nonumber
\eea
where $g$ is the genus of $N_2$, 
\begin{equation}
\int_{N_{2}}\!\!\!\! d^{2}Y \tau R^{(2)} = 8\pi (1-g).  \label{genus}
\end{equation}
This requires the topology of a compact 2-dimensional cross-section of an out-going null hypersurface $N_2$ be either a 2-sphere or a torus \cite{friedman93,chruciel94,jacobson95,galloway96,galloway99}.
This is the reason that the equation
(\ref{topological}) is called a \textit{topological constraint equation}.

\noindent
(v) The evolutions of $\rho_{ab}$ and $\pi^{ab}$ in the $\tau$-time are given by
\bea
& & \hspace{-0.1in}
\ {\partial\over \partial \tau}  \rho_{ab} 
= 2 \tau^{-1}\rho_{a c}\rho_{b d}\pi^{cd}
+\partial_{R}\rho_{ab},    \label {Hamilton1} \\
& & \hspace{-0.1in}
\ {\partial\over \partial \tau} \pi^{ab} 
= -2\tau^{-1}\rho_{c d}\pi^{a c}\pi^{b d}  + \partial_{R}\pi^{ab} 
 +{\tau\over 2}\rho^{ac}\rho^{b  d} (\partial_{R}^{2}\rho_{c d}) \nonumber \\
& & \hspace{0.52in}
-{\tau\over 2}\rho^{a i}\rho^{b j}\rho^{c k} 
(\partial_{R}\rho_{i c})(\partial_{R}\rho_{j k})  .        \label{Hamilton2}
\eea
Notice that the right hand sides of these equations are expressed in terms of $\tau$, $\rho_{ab}$, $\pi^{ab}$, and their $R$-derivatives only. The equations (\ref{Hamilton1}) and (\ref{Hamilton2})
determine the $\tau$ and $R$ dependence of the conformal two-metric and its conjugate.

\noindent
(vi) The evolutions of  $\pi_a$ and $\pi_\tau$ in the $\tau$-time are given by
\bea
& & \hspace{-0.1in}
{\partial \pi_{a}\over \partial \tau}
= 2\tau^{-1}\pi_{a} +(\pi^{bc} + {1\over 2}\rho^{bd}\rho^{ce}\partial_{R}\rho_{de})
{\partial \over \partial Y^{a}} \rho_{bc} \label{aevolution} \nonumber\\
& & \hspace{0.3in}
- {\partial \over \partial Y^{b}}  (2\pi^{bc}\rho_{ac}+\rho^{bc}\partial_{R}\rho_{ac}), \\
& & \hspace{-0.1in}
{\partial \pi_{\tau}\over \partial \tau}= {1\over 2}\tau^{-2} 
+ \tau^{-2}\rho_{ab}\rho_{cd}\pi^{ac}\pi^{bd} 
-2\tau^{-2}{\partial \over \partial Y^{a}}(h\rho^{ab}\pi_{b})   \nonumber\\
& & \hspace{0.3in}
-{1\over 4}\rho^{ab}\rho^{cd}(\partial_{R}\rho_{ac})(\partial_{R}\rho_{bd}). \label{tauevolution}
\eea
The equations (\ref{aevolution}) and (\ref{tauevolution}) are {\it redundant} 
equations, as they can be
obtained by taking the $\tau$ derivatives of $\pi_{\tau}$ and $\tau^{-1}\pi_{a}$ given by
(\ref{pitau}) and (\ref{angko}), using the evolution equations of $\rho_{ab}$ and $\pi^{ab}$ 
given by (\ref{Hamilton1}) and (\ref{Hamilton2}).

Let us outline the procedure of generating solutions to the Einstein's equations 
given by the group of equations (i), (ii), $\cdots$, (vi). 
First, we will solve the evolution equations (\ref{Hamilton1}) and (\ref{Hamilton2})
using ansatz for $\rho_{ab}$ and $\pi^{ab}$. Then we solve the topological constraint 
equation (\ref{topological}) in order to determine 
the $Y^{a}$ dependence of $\rho_{ab}$ and $\pi^{ab}$. 
Then, the superpotential $\ln (-h)$ will be determined by the first-order differential equations (\ref{sp1}), (\ref{sp2}), and (\ref{sp3}). 
We will also calculate the local Hamiltonian density $-\pi_{\tau}$ and the local momentum densities $\pi_{R}$ and $\tau^{-1}\pi_{a}$ using the equations (\ref{pitau}), (\ref{radi}), and (\ref{angko}). This determines the spacetime metric (\ref{ohyoonmetric}) in the privileged coordinates $(\tau, R, Y^{a})$ completely.



\section{Einstein-Rosen Gravitational Waves}
As the first example of solving the Einstein's equations in the privileged coordinates, let us make the following ansatz for $\rho_{ab}$ and $\pi^{ab}$,
\bea
\rho_{ab} = 
\left( \begin{array}{cc}
 f &  0  \\
  0 & 1/f \end{array} \right),
\quad
\pi^{ab} =
\left( \begin{array}{cc}
  \pi_F/f &  0  \\
  0 & - f \pi_F  \end{array} \right)
  \label{ansatz}
\eea
where $f$ and $\pi_F$ are functions of $\tau$ and $R$ to be determined. Namely, we are now assuming two commuting spacelike Killing vectors $\frac{\partial}{\partial Y^1}$ and $\frac{\partial}{\partial Y^2}$ for this ansatz.
Notice that $\pi^{ab}$ is traceless, $\rho_{ab} \pi^{ab} =0$. Substitution of (\ref{ansatz}) into the evolution equations (\ref{Hamilton1}) and (\ref{Hamilton2}) yields the following two equations,
\bea
& & \partial_\tau F  = \frac{4}{\tau} \pi_F + \partial_R F, \label{FH}   \\
& & \partial_\tau \pi_F =  \frac{\tau}{4} \partial_R^2 F + \partial_R \pi_F ,  \label{piH}
\eea
where $F:= 2 \ln f$. The equations (\ref{angko}) and (\ref{sp3}) are trivially satisfied 
due to the Killing condition, and the superpotential equations (\ref{sp1}) and (\ref{sp2}) become, 
\bea
& & 
\partial_\tau \ln (-h) = -\frac{1}{2\tau} + \frac{2}{\tau} \pi_F^2 
+ \frac{\tau}{8}(\partial_R F )^2  \nonumber\\
& & \hspace{0.75in}
+ \pi_F (\partial_R F ), \label{htau}\\
& & 
\partial_R \ln (-h) = \pi_F (\partial_R F). \label{hR}
\eea
By the equation (\ref{FH}), $\pi_F$ can be expressed in terms of $F$,
\bea
\pi_F = \frac{\tau}{4} (\partial_\tau - \partial_R) F.    \label{pif}
\eea
If we plug $\pi_F$ given by the equation (\ref{pif}) into the equations (\ref{piH}), (\ref{htau}), and (\ref{hR}), then they become
\bea
&& \partial_\tau^2 F +\frac{1}{\tau} (\partial_\tau - \partial_R ) F   
- 2 \partial_\tau \partial_R F = 0, \label{main1} \\
&& \partial_\tau \ln(-h) = -\frac{1}{2\tau} + \frac{\tau}{8}(\partial_\tau F)^2, \label{main2}  \\
&& \partial_R \ln(-h) = \frac{\tau}{4} (\partial_\tau F)(\partial_R F) 
- \frac{\tau}{4} (\partial_R F)^2 . \label{main3}
\eea
The equation (\ref{main1}) is a linear second-order PDE of $F$, 
the existence of whose solutions is well-established. Then the metric component $h$ will be determined by the first-order differential equations (\ref{main2}) and (\ref{main3}).
The local Hamiltonian and momentum densities are given by
the constraint equations (\ref{pitau}) and (\ref{radi}), which become,
\bea
&& -\pi_\tau = \frac{1}{2 \tau} 
+ \frac{2}{\tau} ( \pi_F - \frac{\tau}{4} \partial_R F )^2 , \\
&& \pi_R = -\pi_F(\partial_R F). 
\eea
Thus, all the components of the metric tensor are completely determined by a single function $F$ for this ansatz. In fact, this function $F$ is the analog of the news function of the cylindrically symmetric radiative spacetime proposed by H. Bondi et al \cite{Bondi}. All the remaining Einstein's equations (\ref{topological}), (\ref{aevolution}), and (\ref{tauevolution}) are automatically satisfied.

In order to compare this class of spacetimes
with the known solutions to the Einstein's equations, let us make the following coordinate transformation,
\bea
t= \tau + R , \quad \rho = \tau , \quad \phi = Y^1, \quad z = Y^2.  
\eea 
Let us introduce two complex functions $\Psi(t, \rho)$ and $\Gamma(t, \rho)$ defined by
\bea
\Psi = - \ln (f/\rho) + i\pi ,~~~~~ \Gamma = \ln (-2\rho h /f) +i\pi,
\eea
where the imaginary constants were introduced to make our sign convention $h<0$ consistent. 
In the coordinates $(t, \rho, \phi, z)$, the line element (\ref{ohyoonmetric}) becomes
\bea
ds^2 = e^{\Gamma - \Psi} (dt^2 -d\rho^2) 
-\rho^2 e^{-\Psi} d\phi^2 - e^\Psi dz^2,         \label{rosen}
\eea
and the equations (\ref{main1}), (\ref{main2}), and (\ref{main3}), which one must solve
in order to determine the spacetime, become
\bea
&& \partial_t^2 \Psi -\frac{1}{\rho} \partial_\rho \Psi - \partial_\rho^2 \Psi = 0 , \label{si}\\
&& \partial_t \Gamma = \rho (\partial_t \Psi )(\partial_\rho \Psi ), \label{gamma1}\\
&& \partial_\rho \Gamma = \frac{\rho}{2} \{ (\partial_t \Psi )^2 
+ (\partial_\rho \Psi )^2 \}.                \label{gamma2}
\eea
The spacetime (\ref{rosen}) with $\Psi$ and $\Gamma$ determined by (\ref{si}), (\ref{gamma1}), and (\ref{gamma2}) is known as the Einstein-Rosen spacetime
of cylindrically symmetric graviational waves \cite{kuchar71,ER,husain94,brown95,romano96,husain12}. 
This proves that the spacetime generated by solving the Hamilton's equations in this section is precisely the Einstein-Rosen spacetime.

\section{The Schwarzschild Spacetime}
It is well-known that the Schwarzschild solution can be expressed in the form of the canonical Weyl metric. 
We will use the same ansatz (\ref{ansatz}) as in the previous section, and furthermore, 
assume that 
the functions $f$, $\pi_F$, and $h$ depend on $\tau$ and $R$ only as before. If we make 
the following {\it complex} coordinate transformations from $(\tau, R, Y^1, Y^2)$ to $(t, \rho, \phi,z)$ defined by
\bea
t=Y^1,~~~\rho = \tau, ~~~ \phi= i Y^2,~~~z=i(\tau + R), \label{ct}
\eea
and introduce two functions $\psi(\rho,z)$ and $\gamma(\rho,z)$ defined by
\bea
\psi = \frac{1}{2}\ln (\rho f),~~~ \gamma = \frac{1}{2}\ln (-2\rho f h) , \label{pg}
\eea
the line element (\ref{ohyoonmetric}) becomes 
\bea
ds^2 = e^{2\psi} dt^2 - e^{2(\gamma-\psi)} (d\rho^2 + dz^2 ) -\rho^2 e^{-2\psi} d\phi^2, 
\label{Weyl}
\eea
which is just the canonical Weyl metric. In the coordinates $(t, \rho, \phi, z)$, 
the equations  (\ref{main1}), (\ref{main2}), and (\ref{main3}) become
\bea
&&\partial_\rho^2 \psi +\frac{1}{\rho} \partial_\rho \psi + \partial_z^2 \psi = 0 , \\
&&\partial_\rho \gamma = \rho \{ (\partial_\rho \psi )^2 -(\partial_z \psi )^2 \}, \\
&&\partial_z \gamma = 2 \rho (\partial_\rho \psi )(\partial_z \psi ), 
\eea
and the remaining Einstein's equations (\ref{topological}), (\ref{aevolution}), and (\ref{tauevolution}) hold trivially as before.
As is well-known, the canonical Weyl metric represents the Schwarzschild solution 
if $\psi$ and $\gamma$ are given by
\bea
& & \psi = \frac{1}{2} \ln \frac{r_+ +r_- -2m}{r_+ + r_- +2m}, \label{psi}\\
& & \gamma = \frac{1}{2} \ln \frac{(r_+ + r_- )^2 -4m^2}{4r_+ r_-}, \label{gamma} 
\eea
where $r_\pm = [\rho^2 + (z\pm m)^2 ]^{\frac{1}{2}}$ and $m$ is 
a constant\cite{griffith09,misner60,brill63,zipoy66}. 
Moreover, if we make the following coordinate transformations
\bea
\rho = \sqrt{r^2 - 2 m r} \sin \theta , \quad z = (r- m) \cos \theta, 
\label{standardtoweyl}
\eea
then it is easy to show that the line element (\ref{Weyl}) 
becomes the standard line element of the Schwarzschild spacetime
\bea
& &
ds^2 =  (1-\frac{2m}{r}) dt^2 -  (1-\frac{2m}{r})^{-1} dr^2      \nonumber\\
& & \hspace{0.35in}
- r^2 (d\theta^2 + \sin^2{\theta} ~ d\phi^2).   \label{schwarzschild}
\eea
Thus, the solution generated by our method in the privileged coordinates is just 
the Schwarzschild solution in disguise.

It may be instructive to write down explicitly the Schwarzschild solution in the privileged coordinates and its local Hamiltonian and momentum densities.
It is a straightforward exercise to show that the Schwarzschild solution in the privileged coordinates becomes
\bea
ds^2 = -2h~(2dR d\tau + dR^2) +\tau \rho_{ab} ~dY^a dY^b ~,\label{child}
\eea
where 
\bea
\hspace{-0.2in}
\rho_{ab} = 
\left( \begin{array}{cc}
 f &  0  \\
  0 & 1/f \end{array} \right),
\eea
\bea
\hspace{-0.2in}
f= \frac{1}{\tau} \frac{\sqrt{\xi+\sqrt{\xi^2+\eta^2}}
-\sqrt{2}m}{\sqrt{\xi+\sqrt{\xi^2+\eta^2}}+\sqrt{2}m}~,                \label{metric2}
\eea
\bea
\hspace{-0.2in}
h = -\frac{1}{4\sqrt{\xi^2 + \eta^2}} \big( \sqrt{\xi+\sqrt{\xi^2+\eta^2}}+\sqrt{2}m {\big)}^2 ~,    \label{Sch}
\eea
and $\xi$ and $\eta$ are functions of $\tau$ and $R$
given by
\bea
\xi = m^2 -2R\tau -R^2, \quad \eta = 2m(\tau+R)~, \label{xieta}
\eea
respectively.
The local Hamiltonian and momentum densities of the Schwarzschild spacetime can be obtained
by inserting (\ref{metric2}) and (\ref{Sch}) into the equations (\ref{pitau}), (\ref{radi}), and (\ref{angko}),
respectively. They are found to be
\begin{widetext}
\bea
& & \hspace{-0.4in}
-\pi_\tau = \frac{1}{2\tau} 
+ \frac{\tau}{2} \Big[ \frac{2\sqrt{2 }m \{ (m^2 + R^2) (R+2\tau) -R \sqrt{\xi^2 + \eta^2}~ \}} {\sqrt{\xi^2 + \eta^2} \sqrt{\xi+\sqrt{\xi^2 + \eta^2}}(\xi+\sqrt{\xi^2 + \eta^2} - 2m^2)} -\frac{1}{\tau} \Big]^2   
 +\frac{4(\tau+R)}{\sqrt{\xi^2+\eta^2}}\Big( \frac{ \sqrt{\xi+\sqrt{\xi^2 + \eta^2}} -\sqrt{2} m }{\sqrt{\xi+\sqrt{\xi^2 + \eta^2}}}    \nonumber\\
& &
+\frac{R^2 +2R \tau +m^2}{\sqrt{\xi^2 + \eta^2}}  \Big), \label{SchH} \\
& & \hspace{-0.4in}
\pi_R = \frac{2(\tau +R)}{\sqrt{\xi^2 + \eta^2}} \Big( \frac{ \sqrt{\xi+\sqrt{\xi^2 + \eta^2}} -\sqrt{2} m }{\sqrt{\xi+\sqrt{\xi^2 + \eta^2}}}        
+\frac{R^2 +2R \tau +m^2}{\sqrt{\xi^2 + \eta^2}}  \Big) ,  \label{SchR} \\
& & \hspace{-0.4in}
\pi_a = 0. \label{Scha}
\eea 
\end{widetext}
It can be easily shown that the local Hamiltonian and momentum densities given by 
(\ref{SchH}), (\ref{SchR}), and (\ref{Scha}) can be also obtained by taking the gradient of the superpotential $\ln (-h) $ given by (\ref{Sch}), using (\ref{sp1}),  (\ref{sp2}), and (\ref{sp3}). It is clear that both 
$ -\pi_\tau$ and  $\pi_R$ depend on $\tau$ explicitly. This simply reflects the fact that the area time $\tau$ is a geometric object that is quite different from the usual Killing time.

Let us investigate the Schwarzschild solution in the privileged coordinates in the limit
as $m$ approaches zero. In this limit, we find that
\bea
& &\xi \longrightarrow -2 \tau R - R^2, \ \ \eta \longrightarrow 0, \\
& & f \longrightarrow \frac{1}{\tau}, \ \ \ h \longrightarrow -\frac{1}{2},
\eea
respectively. Thus, in the zero mass limit, the line element (\ref{child}) becomes 
\bea
ds^2 = -2dR d\tau - dR^2 + (dY^1)^2  + \tau^2 (dY^2)^2.          \label{flat}
\eea
If we make the following coordinate transformations
\bea
T = R+\tau, \ \ Z =  \tau \cos Y^2, \ \ X =  \tau \sin Y^2,  \ \ Y = Y^{1} ,
\eea
the metric (\ref{flat}) becomes the metric of the Minkowski spacetime 
\bea
ds^2 = -dT^2 + dX^2 + dY^2 + dZ^2 .
\eea
Thus, the metric (\ref{Sch}) becomes the Minkowski metric in the limit $m \longrightarrow 0$, 
as it must be. In this limit, the local Hamiltonian and momentum densities (\ref{SchH}), (\ref{SchR}), and (\ref{Scha}) become
\bea
-\pi_\tau = \frac{1}{\tau}, ~~~~~ \pi_R =  \pi_a =0,
\eea
which agree with the previous results in the paper \cite{Yoon3}. 
Of course, the Hamiltonian density of the Minkowski spacetime can be made zero 
by adding a total derivative term $ - \partial_{\tau} \ln \tau $ to the local
Hamiltonian density $-\pi_\tau$.

Let us examine the event horizon of Schwarzschild spacetime in the privileged coordinates.
Since the event horizon of Schwarzschild spacetime in the standard coordinates is defined by the surface 
$r=2m$, we have 
\bea
\rho = 0,~~~  0 \leq |z| \leq m.
\eea
at the horizon. From the coordinate transformations (\ref{ct}), we find that the horizon of the Schwarzschild spacetime in the privileged coordinates is the region 
\bea
\tau = 0,~~~ 0 \leq |R| \leq m ,
\eea
where the local Hamiltonian and momentum densities are given by
\bea
-\pi_\tau \longrightarrow \frac{1}{\tau}, ~~~ \pi_R \longrightarrow \frac{2R}{m^2 + R^2}, ~~~ \pi_a = 0, 
\eea
respectively. The physical significances of these quantities on the event horizon are yet to be investigated.

\section{Discussion}
In this paper, we have shown that from the solution-generating method based on the Hamiltonian reduction in which the physical degrees of freedom of gravitational fields are 
{\it manifestly isolated} in the privileged coordinates, we successfully generated the Einstein-Rosen and 
the Schwarzschild spacetimes. This method of finding exact solutions to the Einstein's equations is {\it minimal} in the sense that given the conformal two-metric and its conjugate as solutions to the Hamilton's equations of motion and the topological constraint equations, the remaining metric components are determined by the first-order differential equations, known as the superpotential equations. All the other equations
are trivially satisfied, including the constraint equations. Of course, one should be able to find different classes of spacetimes using this solution-generating method, and this problem is under investigation.

We would like to emphasize the importance of the (2+2) Hamiltonian reduction, which has
the following remarkable merits. 
First of all, locally defined Hamiltonian and momentum densities of gravitational fields can be straightforwardly calculated, provided that the metric of the gravitating system is known in the privileged coordinates. The very fact that local Hamiltonian and momentum densities can be defined for generic gravitational fields is a direct consequence of Hamiltonian reduction.
The complete deparametrization of the Einstein's theory is achieved by choosing certain functions of the gravitational phase space as the privileged coordinates in which all the constraints become {\it trivial}. Of course, if one prefers to work in arbitrary coordinates with the spacetime diffeomorphism intact, then the Hamiltonian and momentum densities
of gravitational fields become pure constraints again. After the Hamiltonian reduction, however, all the physical quantities including local Hamiltonian and momentum densities are calculated in the privileged coordinates, and the issue of general covariance is removed
thereby. 

Secondly, in numerically generated spacetimes, it is important to make sure that the constraint equations be satisfied at all later times, given that the constraint equations are satisfied initially. In our Hamiltonian reduction, however, this issue of stability of the constraint equations does not appear. One can bypass the {\it notorious} problem of solving the constraint equations because they are simply the defining equations of the local Hamiltonian and momentum densities of gravitational fields.

Finally, it must be emphasized that we limited our discussions to the case where the 
two-dimensional shift vector vanishes in this paper. Geometrically, the two-dimensional shift is merely a gauge connection associated with the diffeomorphisms of a two-surface.  Therefore it is always  possible to find a \textit{local} coordinate system with a vanishing two-dimensional shift on a hypersurface $\Sigma_\tau$ defined by $\tau =$ constant. The privileged coordinates in our formalism are one of such coordinate charts.
When the topology of $\Sigma_\tau$ is $S^3$, however, such a local coordinate chart with a vanishing  two-dimensional shift can not cover the entire manifold. 
In this case, we must generalize our reduction formalism to include a non-vanishing 
two-dimensional shift. Thus, two-dimensional shift may be regarded as  topological degrees of freedom describing the topology of $\Sigma_\tau$, independent of the dynamical degrees of freedom of gravitational fields \cite{fischer97,fischer00}. It is an interesting problem to find the (2+2) Hamiltonian reduction formalism of the Einstein's theory with a non-vanishing two-dimensional shift. This problem is also under investigation.\\

\section{Acknowledgement} \ \ This work was supported by Konkuk University (2013-A019-0133).

\end{document}